\documentclass[fleqn,usenatbib,useAMS]{aa}  
\usepackage{xspace}

\usepackage{lmodern}
\usepackage{amsmath}    
\usepackage{amssymb}
\usepackage{hyperref}
\usepackage{graphicx}
\usepackage{inputenc}
\usepackage[T1]{fontenc}
\usepackage{epsfig}
\usepackage{epstopdf}
\usepackage{natbib}
\usepackage{relsize}
\usepackage{ae,aecompl}
\usepackage{graphicx}   
\usepackage{times}
\usepackage{subcaption}
\def\lya{Ly$\alpha$}

\def\fesc{$f_\mathrm{esc}$}
\def\fescly{$f_\mathrm{esc}^{Ly\alpha}$~}
\def\fesclyp{$f_\mathrm{esc}^{Ly\alpha}$}
\def\hi{\ion{H}{i}}

\def\ha{H$\alpha$}

\def\nh{$N_\mathrm{H}$}

\newcommand{\oi}{\ion{O}{i}~}

\newcommand{\siii}{\ion{Si}{ii}~}

\newcommand{\oip}{\ion{O}{i}}
\newcommand{\ovip}{\ion{O}{vi}}

\newcommand{\ciip}{\ion{C}{ii}}
\newcommand{\siiip}{\ion{Si}{ii}}

\newcommand{\oiiip}{\ion{O}{iii}}

\newcommand{\oiip}{\ion{O}{ii}}

\newcommand{\nhi}{\ifmmode N_\text{HI} \else $N_\text{HI}$\fi}
\newcommand{\cf}{$C_f$}
\newcommand{\cfhi}{$C_f^\text{H}$}
\newcommand{\cfhip}{$C_f^\text{pre, H}$}
\newcommand{\cfsi}{$C_f^\text{Si}$}
\newcommand{\fesco}{$f_\text{esc}^\text{obs}$}
\newcommand{\fescdb}{$f_\text{esc}^\text{NH}$}
\newcommand{\fescpf}{$f_\text{esc}^\text{CF}$}
\newcommand{\fescd}{$f_\text{esc}^\text{D}$}
\newcommand{\fesct}{$f_\text{esc}^\text{pre}$}
\newcommand{\fesctla}{$f_\text{esc}^\text{pre, Ly$\alpha$}$}
\newcommand{\fesctsi}{$f_\text{esc}^\text{pre, Si}$}
\newcommand{\fescto}{$f_\text{esc}^\text{pre, O}$}
\newcommand{\ott}{O$_{32}$}
\newcommand{\megasaura}{M\textsc{eg}a\textsc{S}a\textsc{ura}}
\newcommand{\ebv}{E$_{B-V}$}
\defcitealias{gazagnes}{Paper~I}
\newcommand{\oh}{\ifmmode 12 + \log({\rm O/H}) \else$12 + \log({\rm O/H})$\fi}

\begin{document}

   \title{Accurately predicting the escape fraction of ionizing photons using rest-frame ultraviolet absorption lines 
   }

 \author{J. Chisholm\inst{1},
S. Gazagnes\inst{1,2,3,4},
D. Schaerer\inst{1,5}, 
A. Verhamme\inst{1}, 
J. R. Rigby\inst{6}, 
M. Bayliss\inst{7}, 
K. Sharon\inst{8}, M. Gladders\inst{9,10}, H. Dahle\inst{11}}
 \offprints{John.Chisholm@unige.ch}
  \institute{
Observatoire de Gen\`eve, Universit\'e de Gen\`eve, 51 Ch. des Maillettes, 1290 Versoix, Switzerland
         \and
         Johan Bernouilli Institute, University of Groningen, P.O Box 407, 9700 Groningen, AK, The Netherlands 
   \and
   Kapteyn Astronomical Institute, University of Groningen, P.O Box 800, 9700 AV Groningen, The Netherlands
   \and
   KVI-Center for Advanced Radiation Technology (KVI-CART), University of Groningen, Zernikelaan 25, Groningen 9747 AA, The Netherlands
         \and
 CNRS, IRAP, 14 Avenue E. Belin, 31400 Toulouse, France
 \and 
 Observational Cosmology Lab, NASA Goddard Space Flight Center, 8800 Greenbelt Rd., Greenbelt, MD 20771, USA
 \and 
 MIT Kavli Institute for Astrophysics and Space Research, 77 Massachusetts Ave., Cambridge, MA 02139, USA
 \and 
 Department of Astronomy, University of Michigan, 500 Church St., Ann Arbor, MI 48109, USA
 \and 
 Department of Astronomy \& Astrophysics, University of Chicago, 5640 S. Ellis Ave., Chicago, IL 60637, USA
 \and 
Kavli Institute for Cosmological Physics, University of Chicago, 5640 South Ellis Ave., Chicago, IL 60637, USA
\and 
Institute of Theoretical Astrophysics, University of Oslo, P.O. Box 1029, Blindern, NO-0315 Oslo, Norway}

\authorrunning{Chisholm et al.}
\titlerunning{Accurately predicting the escape fraction of ionizing photons}

\date{Received date; accepted date}

 
  \abstract
    {The fraction of ionizing photons that escape high-redshift galaxies sensitively determines whether galaxies reionized the early universe. However, this escape fraction cannot be measured from high-redshift galaxies because the opacity of the intergalactic medium is large at high redshifts. Without methods to  {indirectly} measure the escape fraction of high-redshift galaxies, it is unlikely that we will know what reionized the universe. Here, we analyze the far-ultraviolet (UV) \ion{H}{i} (Lyman series) and low-ionization metal absorption lines of nine low-redshift, confirmed Lyman continuum emitting galaxies. We use the \ion{H}{i} covering fractions, column densities, and  dust attenuations measured in a companion paper to predict the escape fraction of ionizing photons. We find good agreement between the predicted and observed Lyman continuum escape fractions (within $1.4\sigma$) using both the \hi\  and  {ISM} absorption lines. The ionizing photons escape through holes in the \hi, but we show that dust attenuation reduces the fraction of photons that escape galaxies. This means that the average high-redshift galaxy likely emits more ionizing photons than low-redshift galaxies. Two other indirect methods accurately predict the escape fractions: the \lya\ escape fraction and the optical  [\ion{O}{iii}]/[\ion{O}{ii}] flux ratio. We use these indirect methods to predict the escape fraction of a sample of 21 galaxies with rest-frame UV spectra but without Lyman continuum observations. Many of these galaxies have low escape fractions (\fesc~$\le 1$\%),  but 11 have escape fractions $>1$\%. The methods presented here  {will} measure the escape fractions of high-redshift galaxies, enabling future telescopes to determine whether star-forming galaxies reionized the early universe. }


   \maketitle
%
\section{Introduction}
In the local universe, gas between galaxies is mostly highly ionized \citep{fan2006}, but it has not always been that way. Hydrogen recombined at $z=1090$ and remained neutral until $z\sim7-9$ \citep{planck}. This is most easily observed by the absorption blueward of rest-frame Ly$\alpha$ (1216\AA) in the spectra of $z > 6$ quasars \citep[the "Gunn-Peterson trough";][]{gunn, becker}. Some mechanism must have produced copious ionizing photons to reionize the universe.

The source of reionization is one of the core questions that future large observatories, such as the {\it James Webb} Space Telescope ({\it JWST}) and extremely large telescopes (ELT), aim to answer. One possibility is that active galactic nuclei (AGN) provided the ionizing photons. However, current observed AGN luminosity functions indicate that there were not enough AGN to reionize the early universe \citep{hopkins08, willott, Fontanot12, ricci, onoue}.
\\

An alternative source of ionizing photons is the first generation of high-mass stars. For these stars to matter to reionization, the emissivity of ionizing photons  ($\dot{n}_\text{ion}$) escaping high-redshift galaxies must exceed the recombination rate. Commonly $\dot{n}_\text{ion}$ is expressed as
\begin{equation}
    \dot{n}_\text{ion} = f_\text{esc} \xi_\text{ion} \rho_\text{UV},
    \label{eq:emis}
\end{equation}
where $\xi_\text{ion}$ is the intrinsic number of ionizing photons emitted by stars, $\rho_\text{UV}$ is the total  ultraviolet (UV) luminosity density, and \fesc\ is the a{absolute} fraction of ionizing photons that escape galaxies. More generally, the quantities in Eq.~\ref{eq:emis} depend on the UV magnitude, $M_\text{UV}$, and the total $\dot{n}_\text{ion}$ is found by integrating over the UV luminosity function.  While highly dependent on clumping and redshift, the estimated $\Omega_\text{matter}$ from $\Lambda$CDM indicates that the universe is reionized when log($\dot{n}_\text{ion}$[photons~s$^{-1}$~Mpc$^{-3}$]) is near $50-51$ \citep{madau99, Meiksin, bolton, Ouchi09, Kuhlen, robertson13, robertson15}. 

In principle, whether or not stars reionized the universe is an observable question. The parameter $\xi_\text{ion}$ is related to the observed H$\alpha$ emission and depends on the metallicity and star formation rate of the galaxies \citep{leitherer95, bruzual}. Recent studies constrain $\xi_\text{ion}$ at $z = 6-8$ \citep{dunlop12, bouwens12, robertson13, harikane}. Similarly, deep {\it Hubble} Space Telescope (HST) observations have pushed the UV luminosity functions down to fainter $M_\text{UV}$ at high redshifts \citep{bouwens06, Ouchi09, oesch14, finkelstein15, bouwens15, livermore17, oesch17}. While requiring extraordinary observations, these studies are beginning to constrain $\xi_\text{ion}$ and $\rho_\text{UV}$ during the epoch of reionization.

These observational constraints suggest that \fesc$~\sim0.1-0.2$ if stars reionized the universe \citep{Ouchi09, robertson13, bouwens15, dressler15, robertson15, ishigaki}. Whether \fesc\ reaches these values has not been observationally confirmed. First, the opacity of the intergalactic medium (IGM) is, on average, too large to observe {LyC photons above $z\sim4$ \citep{worseck14}}. Therefore, a direct detection of ionizing photons escaping from a single galaxy during the epoch of reionization is statistically unlikely. Alternatively, studies focused on lower redshift galaxies where the Lyman continuum (LyC; $<912$\AA) is directly observable. However, directly detecting ionizing photons at low redshift is still challenging. It requires deep observations of intrinsically faint emission in the very far-UV, which is a notoriously hard regime for high-sensitivity detectors. Only ten individual $z < 0.4$ galaxies have spectroscopically confirmed \fesc~$> 0$ \citep{Bergvall06,leitet11, Borthakur14, Izotov16, Izotov16b, Leitherer16, izotov17}. Additionally, four such galaxies at $z\sim3-4$ have been confirmed \citep{Vanzella15, Vanzella16, deBarros16,shapley16, bian, vanzella17}, after accounting for foreground contamination \citep[e.g.,][]{Vanzella10}. To constrain \fesc\ during the epoch of reionization, indirect \fesc\ probes available at both high (to measure galaxies in the epoch of reionziation) and low redshifts (to confirm the predicted \fesc\ values) are required. 

We present a new analysis of the rest-frame UV properties of nine confirmed low-redshift galaxies that emit ionizing photons and have publicly available far-UV observations. We use the fits of the stellar continua, interstellar medium (ISM) metal absorption lines, and ISM  \ion{H}{i} absorption lines (the Lyman series) from \citet{gazagnes} (hereafter Paper~I) to constrain the neutral gas and dust attenuation properties. Since the \hi\ and dust are the major sinks of ionizing photons, these measurements allow us to accurately predict \fesc. These new methods can be used to efficiently select low-redshift galaxies that emit ionizing photons or for future telescopes (such as {\it JWST} or ELTs) to constrain $\dot{n}_\text{ion}$ of galaxies reionizing the universe.

The structure of this paper is as follows: Sect.~\ref{data} introduces the observations of the nine publicly available LyC emitters and summarizes how \citetalias{gazagnes} fit the Lyman series absorption lines. We use these fits to predict \fesc\ (Sect.~\ref{direct}) and explore what fit parameters contribute to the observed \fesc\ values (Sect.~\ref{methods}). We then test using the \siii absorption lines (Sect.~\ref{si2}), \lya\ escape fractions (Sect.~\ref{lya}), and the [\ion{O}{iii}]/[\ion{O}{ii}] ratios (Sect.~\ref{o3}) to indirectly predict \fesc. In Sect.~\ref{horse} we apply these indirect methods to galaxies without Lyman series observations to demonstrate how these methods can be used for high-redshift galaxies. Our main conclusions are summarized in Sect.~\ref{summary}.

\begin{table*}
\centering
\caption{Measured properties of the Lyman continuum emitting sample from \citet{gazagnes} in order of decreasing \fesco.}
\resizebox{\textwidth}{!}{
\begin{tabular}{lccccccccc}
\hline
\hline
Galaxy name & \fesco\  & E$_{B-V}$ & $\log(\nhi)$ & \cfhi & \cfsi& \fesct & \fesclyp{}  & \ott\ \\
 &    & [mag] & [log(cm$^{-2}$)] & & & \\
\hline 
(1) & (2)  & (3) & (4) & (5) & (6) & (7) & (8) & (9) \\
\hline 
J115204.9$+$340050 & $0.13 \pm 0.01$ & $0.13 \pm 0.02$ & $19.43\pm 0.18$ &$0.62\pm0.09$ & $0.27\pm0.14$& $0.08\pm0.02$ &$0.34\pm0.07$ & 5.4\\
J144231.4$-$020952&  $0.074\pm0.010$ &$0.14 \pm 0.02$ & $19.69\pm0.58$ &$0.55 \pm 0.04$  & $0.47\pm0.19$ &  $0.09 \pm 0.02$ & $0.54\pm0.11$ & 6.7\\
J092532.4$+$140313 &  $0.072\pm0.008$ & $0.16\pm0.02$ & $17.81\pm3.0^\text{H}$ & $0.64\pm0.09$& $0.41\pm0.19$ &  $0.05\pm0.01$ & $0.29\pm0.06$  & 4.8\\ 
J150342.8$+$364451 & $0.058\pm0.006$ & $0.27\pm0.04$ & $19.60\pm0.17$ & $0.75\pm0.06$ & $0.45 \pm 0.28$ & $0.010 \pm 0.005$ & $0.29\pm0.06$ & 4.9 \\
J133304.0$+$624604 & $0.056\pm0.015$ & $0.15\pm0.04$ & $19.78\pm0.37$ &$0.83 \pm 0.07$ & $0.39\pm0.21$ & $0.03 \pm 0.01$&$0.52\pm0.11$ & 4.8\\
Tol~0440$-$381  & $0.019\pm0.010$ & $0.27\pm0.03$ & $19.27\pm0.10$ &$0.57\pm0.08$ & $0.37\pm0.05$  & $0.017 \pm 0.006$& -- & 2.0\\
J092159.4$+$450912 & $0.010\pm0.001$  & $0.22\pm0.02$ & $18.63\pm0.19$ &$0.77\pm0.12$ & $0.60\pm0.14$ & $0.017 \pm 0.004$& $0.01\pm0.01$& 0.3 \\
Tol~1247$-$232 & $0.004\pm 0.002$ & $0.16\pm0.01$ & $19.19\pm0.44$& $0.69\pm0.08$ & $0.26 \pm 0.01$ & $0.049\pm 0.008$ &$0.19\pm0.01$ & 3.4\\
Mrk~54 & $<0.002$  & $0.36\pm0.01$ &$19.37\pm0.10$ &$0.50 \pm 0.08$ & $0.32\pm0.01$ & $0.007 \pm 0.002$& --& 0.4 \\
\hline
\end{tabular}
}
\tablefoot{Column 1 gives the name of the galaxy; column 2 gives the observed escape fraction of ionizing photons \citep[\fesco; taken from the recalculations of][]{chisholm17}.  Column 3 is the stellar continuum attenuation (\ebv). Column 4 is the logarithm of the \hi\ column density (\nhi) derived from the \oip~1039\AA\ absorption line and \oh\ (except for J0925$+$1403 where \oi is not detected; denoted with an H). Column 5 is the \hi\ covering fraction (\cfhi) derived from the depth at line center of the Lyman series absorption lines, and column 6 is the \siii covering fraction (\cfsi). Columns 3--6 are taken from \citetalias{gazagnes}. Column 7 is the predicted Lyman continuum escape fraction using the dust attenuation and \cfhi\ (Eq.~\ref{eq:esc}). Column 8 is the \lya\ escape fraction \citep[\fesclyp;][]{verhamme16} rescaled to an intrinsic flux ratio of \lya/\ha~$= 8.7$. The extinction-corrected [\ion{O}{iii}]~5007\AA/[\ion{O}{ii}]~3727\AA\ flux ratio (\ott)  is given in column 9 \citep{verhamme16}.  We note that Tol~0400$-$381 and Mrk~54 have the detector gap over the \lya\ line, thus they do not have a measured \fescly. }
\label{tab:sample}
\end{table*}

\section{Data and absorption line analysis}
\label{data}
\subsection{Rest-frame far-UV observations}
\subsubsection{ {The} Lyman continuum emitting sample}
In this paper, we predominantly use the rest-frame far-UV spectra of the nine publicly available known LyC emitters \citep[hereafter called the Lyman continuum emitting sample;][]{Borthakur14, Izotov16, Izotov16b, Leitherer16} taken with the Cosmic Origins Spectrograph \citep[COS;][]{green12} on the HST. We note that \citet{izotov17} recently discovered a tenth Lyman continuum emitter that we do not include in this paper because it is not publicly available (but see Sect.~\ref{J1154}). As summarized in \citet{chisholm17}, these nine galaxies have low stellar masses (10$^{8}-10^{10}$~M$_\odot$), high star formation rates ($3-77$~M$_\odot$~yr$^{-1}$), and moderately low gas-phase metallicities ($\oh = 7.9-8.7$). Table~\ref{tab:sample} lists the galaxies in the Lyman continuum emitting sample and their observed Lyman continuum \citep[\fesco;][]{chisholm17} and \lya\  \citep[\fesclyp;][]{verhamme16} escape fractions. Two galaxies, Tol~0440$-381$ and Mrk~54, have the COS detector gap over the \lya\ feature. Therefore, their \fescly values are not measured.

Eight of these nine galaxies were observed with the low-resolution G140L grating (nominal resolution of $R\sim1500$) on HST/COS, while J0921$+$4509 was observed with the high-resolution G130M and G160M gratings ($R\sim15000$). These setups observed the rest-frame Lyman series and \siiip~1260\AA\ absorption lines of each galaxy. Each galaxy also has rest-frame optical observations, such that extinction-corrected [\ion{O}{iii}]~5007\AA/[\ion{O}{ii}]~3727\AA\ flux ratios (\ott) are measured \citep[last column of Table~\ref{tab:sample};][]{verhamme16}. 

The HST/COS G140L data were reduced using the methods outlined in \citet{Worseck16}. Special attention was paid to the pulse heights and extraction apertures of each individual spectrum. The pulse heights and apertures used were outlined in \citet{chisholm17}. We placed the galaxy into the rest frame using the redshifts from the Sloan Digital Sky Survey \citep{Ahn14}. We then corrected each spectrum for foreground reddening using the values from \citet{schlegel} and the Milky Way reddening law \citep{Cardelli89}.  

\subsubsection{Low-redshift galaxies with unobserved LyC emission}
\label{othercos}
In Sect.~\ref{o3} we extend the Lyman continuum emitting sample to include the full sample from \citetalias{gazagnes} with measured \ott\ (see Table~\ref{tab:full}).  This full sample includes four low-redshift galaxies that do not have observations of the Lyman continuum, but have observations of the Lyman series. The full sample includes three Green Pea galaxies \citep{Henry15} and one Lyman Break Analog \citep{Heckman11, Heckman15, Alexandroff15, Heckman16}. These four galaxies were also observed with HST/COS and the G130M grating. The data were reduced following the methods outlined in \citet{wakker2015}. These galaxies do not have LyC observations, consequently we predict their LyC escape fractions but we cannot confirm them. In Sect.~\ref{o3} we use the Lyman series observations of the full sample to predict the relation between \fesc\ and \ott.

\subsubsection{High-redshift galaxies from \megasaura}
\label{othermega}
Similarly, in Sect.~\ref{horse} we focus on 14 $z >2$ lensed galaxies from The Magellan Evolution of Galaxies Spectroscopic and Ultraviolet Reference Atlas \citep[\megasaura;][]{rigby17a}. These lensed galaxies have spectra taken with the MagE spectrograph \citep{mage} on the Magellan telescopes. The data were reduced using D. Kelson's pipeline\footnote{\url{http://code.obs.carnegiescience.edu/mage-pipeline}} and placed into the observed frame using the redshifts measured from the UV emission lines \citep{rigby17a}. Two of these galaxies have Lyman series and \ott\ observations, thus they are included in the full sample (Table~\ref{tab:full}). The other 12 galaxies do not have Lyman series or \ott\ observations, and we apply our indirect methods to these spectra in Sect.~\ref{horse}. These high-redshift galaxies do not have measured Lyman continuum escape fractions, but their rest-frame UV spectra test the methods presented in this paper. 

\subsection{Lyman series fitting}
\label{fit}
To predict the fraction of ionizing photons that escape a galaxy, we determined the \hi\ properties from the Lyman series absorption lines between 920-1025\AA. These measurements describe the quantity and porosity of \hi\ along the line of sight. \citetalias{gazagnes} describes this procedure in detail; here we summarize the process and further details are provided in that paper. 

We fit the observed flux density ($F_\lambda^\text{obs}$) using a linear combination of fully theoretical, {\small STARBURST99} stellar continuum models \citep[$F_\lambda^\star$;][]{claus99}. We created these stellar continuum models using the Geneva stellar evolution tracks \citep{geneva94} and the WM-BASIC method \citep{claus2010}, assuming an initial mass function with a high (low) mass exponent of 2.3 (1.3) and a high-mass cutoff at 100~M$_\odot$. These models have a spectral resolution of $R\sim2500$. The final $F_\lambda^\star$ is a linear combination of 10 single-age stellar continuum models each with an age between $1-40$~Myr. The stellar continuum metallicity was chosen as the model closest to the measured gas-phase metallicity. We fit for the linear coefficient multiplied by each single-aged {\small STARBURST99} model that best matches the data using {\small MPFIT} \citep{mpfit}.

We simultaneously reddened $F_\lambda^\star$ to account for a uniform foreground dust screen using the attenuation law ($k_\lambda$) from \citet{reddy_ext} and a fitted stellar attenuation value (\ebv). In Sect.~\ref{geometry} we discuss the implications for the assumed dust geometry.

Finally, we measured the \hi\ and metal ISM absorption line properties by including Lyman series, \ovip, \oip, \ciip, \ion{C}{iii}, and \siii absorption features. We fit for the observed Lyman series absorption lines using the radiative transfer equation, assuming an overlapping covering fraction \citep[\cf;][]{barlow97, hamann97}, which has a functional form of
\begin{equation}
F_{\lambda}^\text{obs} = F_\lambda^\star \times 10^{-0.4 E_\text{B-V} k_\lambda} \times \left( 1-C_f^\text{H} +C_f^\text{H} \text{e}^{-\tau_\lambda}\right),
\label{eq:rad}
\end{equation}
where we fit for \ebv, the intrinsic stellar continuum ($F_\lambda^\star$),  the optical depth ($\tau = \sigma N_\text{HI}$), and the \hi\ covering fraction (\cfhi). As discussed in \citetalias{gazagnes}, the \hi\ lines are saturated ($\tau_\lambda \gg 1$), but not damped. Consequently, \nhi\ cannot be accurately determined. Therefore, we measured the \hi\ column density from the unsaturated \oip~1039\AA\ line, and converted this column density into \nhi\ using the observed \oh. One galaxy, J0925$+$1403, does not have a \oi~1039\AA\ detection, therefore we used the fitted \nhi\ value and the large associated errors. The fits of Eq.~\ref{eq:rad} constrain the stellar population, dust and \nhi\ properties of the LyC emitters. The Lyman series fits for all of the galaxies are shown in the Appendix of \citetalias{gazagnes}. 

Since the Lyman series is always found to be optically thick \citepalias{gazagnes}, we find that \cfhi\ is most robustly measured by taking the median of
\begin{equation}
    C_f^\text{H} = 1 - \frac{F_\lambda^\text{obs}}{F_\lambda^\star 10^{-0.4E_\text{B-V}k_\lambda}}
    \label{eq:cf}
\end{equation}
in a region that we visually selected near each Lyman series line. To calculate the \cfhi\ errors of the individual Lyman series transitions, we varied the observed flux by a Gaussian distribution centered on zero with a standard deviation equal to the flux error. We then measured \cfhi\ from this altered flux array and tabulate the result. We repeated the process 1000 times to produce a distribution of \cfhi\ values. We then took the median and standard deviation of this distribution as the \cfhi\ estimate and uncertainty. After we measured \cfhi\ for each transition,  we took the weighted median and standard deviation of all observed Lyman series lines as the \cfhi\ estimate and error (see Table~\ref{tab:sample}). We used this method because it does not rely on assumptions about how the \cfhi\ changes with velocity, and we could control for the impact of nearby Milky Way absorption lines.

\subsection{\ion{Si}{ii} observations}

Finally, we measured the \siii covering fraction (\cfsi) in two ways. First, we measured \cfsi\ of the \siiip~1260\AA\ line with Eq.~\ref{eq:cf}. This method assumes that the strong \siiip~1260\AA\ line, with an oscillator strength of 1.22, is saturated. Second, we calculated \cfsi\ from the \siiip~1190\AA\ doublet, which accounts for low \siii optical depths. We took the average and standard deviation of these two values as the \cfsi\ values and errors, respectively. We note that both estimates of \cfsi\ are largely equivalent to each other, implying that the \siiip~1260\AA\ line is saturated \citepalias[see][]{gazagnes}. Now we have measured the ingredients to predict the Lyman continuum escape fractions.

\section{Predicting the Lyman continuum escape fraction with the Lyman series}
\label{direct}
\begin{figure}
\includegraphics[width = 0.5\textwidth]{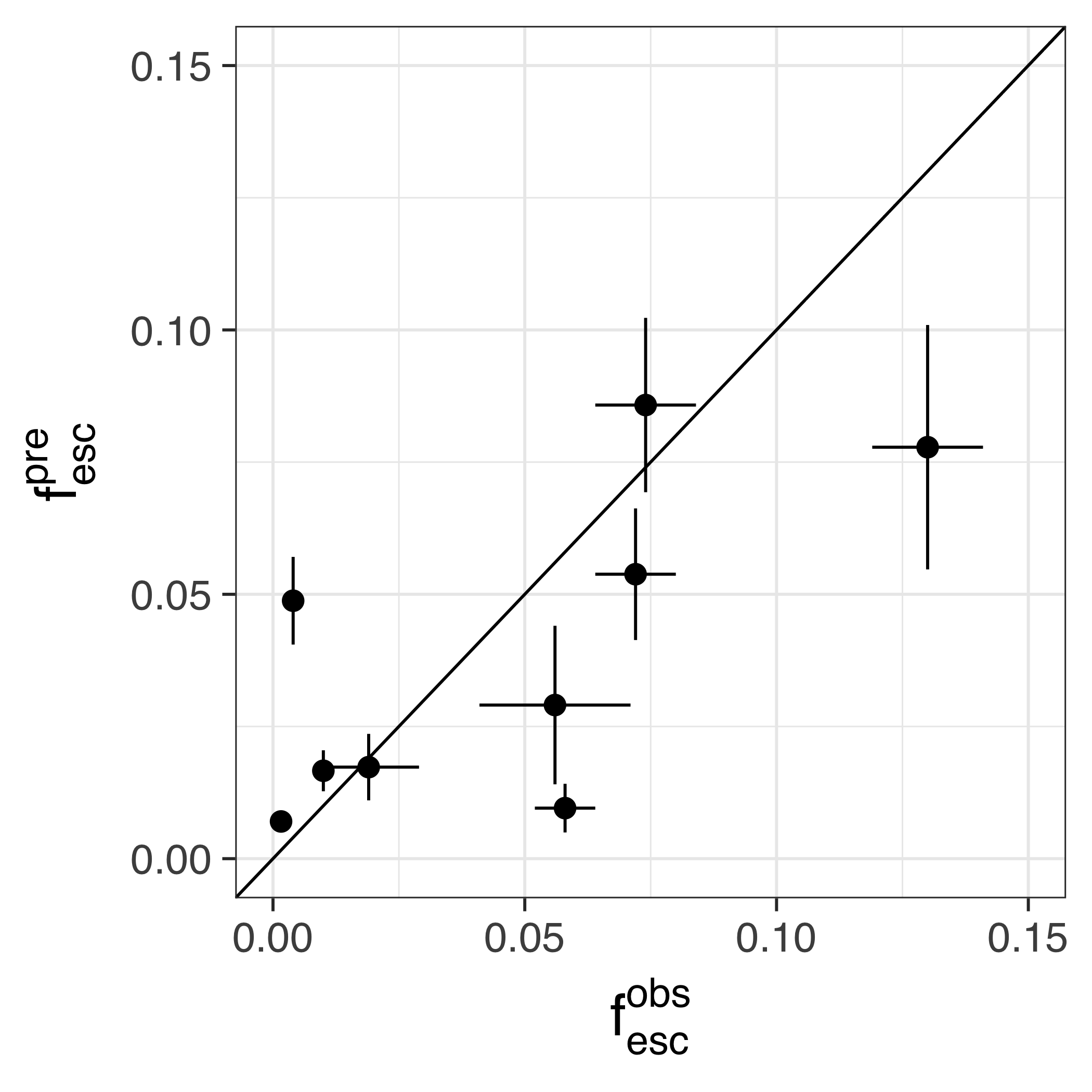}
\caption{Plot of the observed Lyman continuum escape fraction (\fesco) vs. the predicted Lyman continuum escape fraction (\fesct) computed using the observed \hi\ absorption properties and  Eq.~\ref{eq:esc}. The solid line shows a one-to-one relation, indicating that the predicted values are within 1.4$\sigma$ of the observed Lyman continuum escape fractions. We note that there are two outliers more than 3$\sigma$ from the one-to-one relation: Tol~1247-232 (at \fesct~$\sim0.05$ and \fesco~$\sim 0.005$) and J1503$+$3644 (at \fesct~$\sim0.01$ and \fesco~$\sim 0.06$). These outliers are discussed in Sect.~\ref{direct}}.
\label{fig:fesc}
\end{figure}

The absolute Lyman continuum escape fraction, \fesc, is defined as the ratio of the observed ionizing flux to the intrinsic ionizing flux produced by stars,
\begin{equation}
f_\text{esc} = \frac{F^\text{obs}_{912}}{F^{\star}_{912}} ,
\end{equation}
where $F_\lambda^\text{obs}$ is defined in Eq.~\ref{eq:rad}. Since ionizing photons can be absorbed by dust or \hi, \fesc\ is predicted from the fits to the Lyman series and the dust attenuation as
\begin{equation}
    f^\text{pre}_\text{esc} = 10^{-0.4\text{E$_{B-V}$}k_{912}} \times \left(1-C_f^\text{H}\right).
    \label{eq:esc}
\end{equation}
The Lyman continuum is optically thick at \hi\ column densities above 10$^{17.7}$~cm$^{-2}$. For column densities below this column density, the gas is optically thin and the escape fraction increases because unabsorbed light escapes. However, in \citetalias{gazagnes} we used the \oi\ column densities to demonstrate that the \nh\ in these galaxies is larger than 10$^{18.63}$~cm$^{-2}$. Therefore, we neglected the last term of Eq.~\ref{eq:rad}  when calculating \fesct. To calculate \fesct, we used  $k_{912}=12.87$ from the attenuation curve of \citet{reddy_ext}. The errors on \fesct\ were calculated by propagating the errors of \ebv\ and \cfhi\ through Eq.~\ref{eq:esc}. 

The value \fesct\ closely follows \fesco\ for the nine galaxies in the Lyman continuum emitting sample (Fig.~\ref{fig:fesc}). The normalized absolute difference between \fesct\ and \fesco\ (|\fesco-\fesct|/\fesco) is 48\%. The median \fesct\ is within 1.4$\sigma$ of \fesco\ (i.e., within the 95\% confidence interval). This assumes a uniform distribution because the reported \cfhi\ and \ebv\ errors are highly non-Gaussian. The value \fesco\ heavily depends on the modeling of the stellar population. Table~9 of \citet{Izotov16b} demonstrates that the median \fesco\ varies by 0.01 (10-20\%) if different stellar population models are used. This error, while not accounted for in the standard \fesco\ error bars, would improve the quoted statistics.

Two galaxies have \fesct\ more than 3$\sigma$ from \fesco: Tol~1247$-$232 and J1503$+$3644. For Tol~1247$-$232, \fesco\ is challenging to measure because it is a low-redshift galaxy with possible geocoronal \lya\ contamination \citep{chisholm17}. Other studies, which used the same observations but different reductions and handling of geocoronal \lya, have measured \fesco$=0.045\pm0.012$ and $0.015\pm0.005$ \citep[][respectively]{Leitherer16, Puschnig}, whereas \citet{chisholm17} have measured \fesco$=0.004\pm0.002$. These values are more consistent with the derived \fesct$=0.049\pm0.008$. In reality, it is remarkable that \fesct\ and \fesco\ are at all similar. Regardless, we conclude that Eq.~\ref{eq:esc} accurately reproduces the observed LyC escape fractions to within 1.4$\sigma$, on average.  

\subsection{Effect of the assumed geometry on \fesc}
\label{geometry}
The \fesc\ is measured along the line of sight from a star-forming region to the observer and line-of-sight geometric effects could impact \fesc. To estimate \fesct, we assumed a uniform dust screen (Eq.~\ref{eq:esc}). This posits that the dust is uniformly distributed along the line of sight to the galaxy. It is worth exploring the effect this assumed geometry has on \fesct. Detailed discussions on this issue are also provided elsewhere \citep[][]{zackrisson13, vasei, reddy, gazagnes}.  

A simple alternative geometry is that the dust only resides within clumpy neutral gas clouds. Between these neutral clouds are dustless and gasless holes, which we call  {a} clumpy geometry. This geometry alters the radiative transfer equation (Eq.~\ref{eq:rad}) to become
\begin{equation}
    F_\lambda^\text{obs, clumpy} = F_\lambda^\star \times 10^{-0.4 E_\text{B-V} k_\lambda}\times C_f^\text{H} \text{e}^{-\tau_\lambda}+F_\lambda^\star \times (1-C_f^\text{H}), 
    \label{eq:clumpy}
\end{equation} 
and the ionizing escape fraction is
\begin{equation}
     f^\text{pre, clumpy}_\text{esc} = C_f^\text{H} \times 10^{-0.4E_\text{B-V}k_{912}} \times \text{e}^{-\tau_\lambda} +  \left(1-C_f^\text{H}\right).
    \label{eq:dustless}
\end{equation}
 {We note that the clumpy and uniform geometries treat the dust differently. } In  {the} clumpy geometry, the dust attenuation acts only on the $e^{-\tau_\lambda}$ term. To remain at the same $F_\lambda^\text{obs}$ (or \fesc), the \cfhi\ and \ebv\ of  {the} clumpy geometry must be larger than  {the} uniform geometry. This is because unattenuated light passes through holes in clumpy geometry, forcing the attenuation within the clumps to be stronger, and the holes to be smaller, to match the observed flux. 

To test the effect of the geometry, in \citetalias{gazagnes} we refit $F_\lambda^\text{obs}$ from J1152$+$3400 and J0921$+$4509, a large and a small \fesco\ galaxy, with the clumpy model (Eq.~\ref{eq:clumpy}). We find that $C_f^{H}=0.912, 0.976$ and $E_\text{B-V}=0.239$ and $0.236$, respectively. Both are larger than the uniform dust screen model (Table~\ref{tab:sample}). However, these values and Eq.~\ref{eq:dustless} lead to \fesct$=0.088$ and $0.024$, statistically consistent with \fesct\ using the uniform screen (0.08 and 0.016 respectively). 

The fitted values (\ebv, \cfhi) change to match $F_\lambda^\text{obs}$ based on the assumed geometry. Therefore, parameters such as \cfhi\ and \ebv\ are model dependent. However, \fesc\ is model independent because the best combination of the model and the parameters are fit to match the data \citepalias[as discussed in][]{gazagnes}. The geometry must be accounted for --- and remembered --- when comparing and interpreting \cf\ and \ebv, but the \fesc\ values do not strongly depend on the assumed geometry.

\section{Parameters contributing to the predicted escape fractions}
\label{methods}

The previous section showed that the fits to the observed flux accurately predict the escape fraction of ionizing photons. \hi\ column density, \hi\ covering fraction, and dust attenuation determine these fits. The natural question is which parameters contribute to the predicted escape fractions {?} In the next three subsections we explore the contribution of each estimated parameter to the predicted escape fractions. We note that the following analysis does not refit the data to maximize the contribution of each parameter, rather it uses the previous fits to answer which parameters contribute to the predicted escape fractions.

\begin{figure}
\includegraphics[width = 0.5\textwidth]{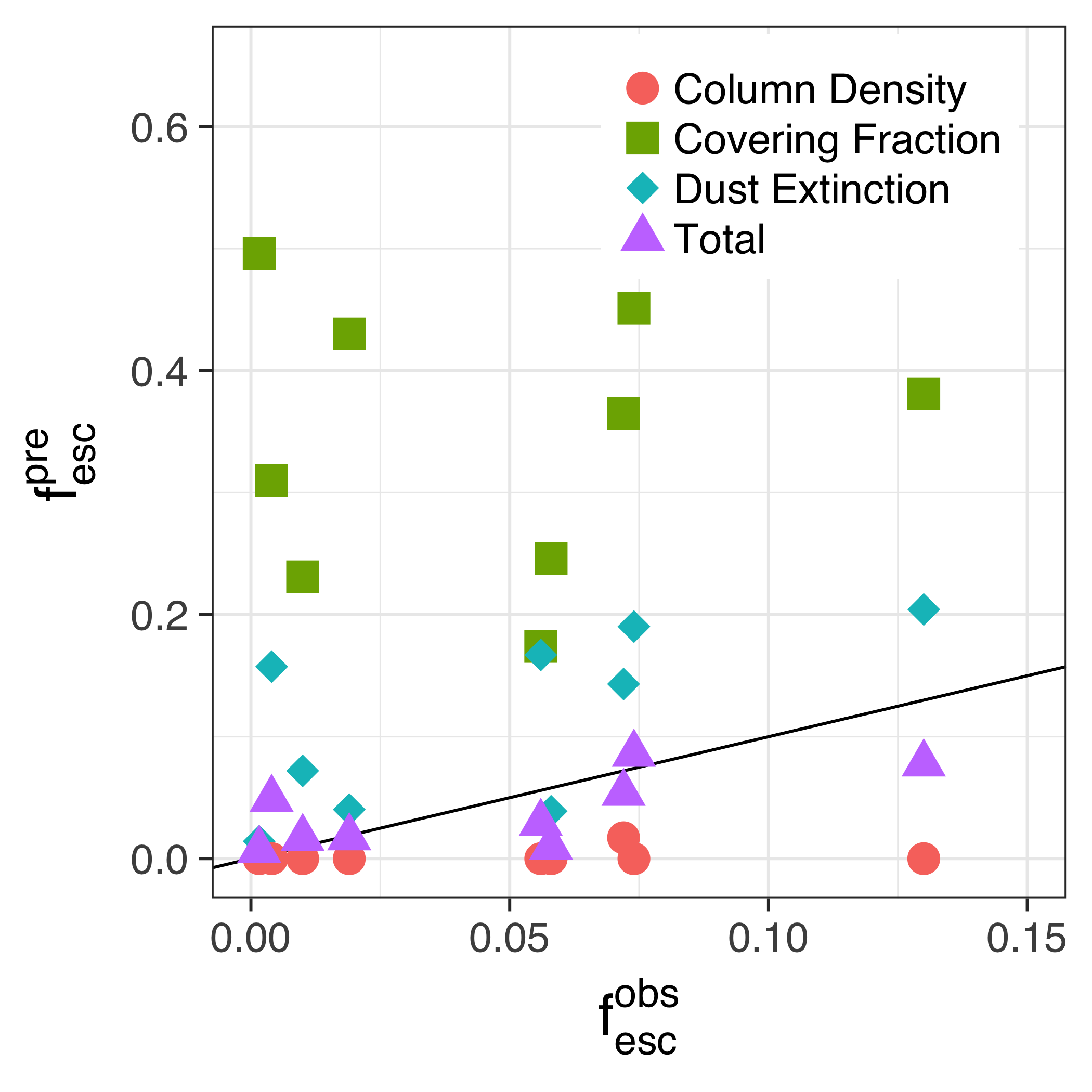}
\caption{Observed Lyman continuum escape fraction (\fesco) vs. the Lyman continuum escape fractions predicted by isolating various fit parameters (\fesct). Each colored symbol represents the contribution of a single parameter from our model (Eq~\ref{eq:esc}). The red circles correspond to the contribution to the escape fraction from the \hi\ column density alone. The cyan diamonds correspond to the contribution from dust attenuation only. The green squares indicate the contribution to \fesct\ from the \hi\ covering fraction. The purple triangles show the combination of all three mechanisms that scatter about the one-to-one line. Dust and the \hi\ covering fraction dominate \fesct.}
\label{fig:methods}
\end{figure}

\subsection{\hi\ column density}
\label{density}
The first parameter that we discuss is \nhi. If \nhi\ is low enough, ionizing photons pass through the ISM unabsorbed \citep[a "density-bounded" region;][]{Jaskot13, zackrisson13, Nakajima14}. The escape fraction of ionizing photons only due to \nhi\ is 
\begin{equation}
    f_\text{esc}^\text{NH} = \text{e}^{-\sigma N_\text{HI}}
,\end{equation}
where $\sigma$ is the photoionization cross section ($6.3\times10^{-18}$~cm$^{2}$). We set \cfhi~=~1 and \ebv~=~0 in Eq.~\ref{eq:esc}. The \fescdb\ values are too low to match \fesco\ (red circles in Fig.~\ref{fig:methods}). This implies that the \hi\ along the line of sight is optically thick \citepalias[see the discussion in][]{gazagnes}. 

\subsection{Covering fraction}
\label{pf}

The second parameter, the covering fraction, implies that ionizing photons escape through holes in the \hi\ gas \citep{Heckman11}. If we assumed no attenuation from dust (\ebv~=~0) and that \hi\ is optically thick, the predicted escape fractions are
\begin{equation}
    f_\text{esc}^\text{CF} = 1 - C_f^\text{H},
\end{equation}
which is greater than 0 for the nine Lyman continuum emitters (green squares in Fig.~\ref{fig:methods}). However, these  \fescpf\ values are substantially higher than \fesco. If holes in the \hi\ were solely responsible for the escape of ionizing photons, and there was no dust, the escape fractions would be much higher than observed. 

Several previous studies have used \fescpf to estimate \fesc, but overestimated the \fesc\ values \citep[][]{quider, Heckman11, Jones12, Jones13, leethochawalit, vasei}. For example, \citet{quider} obtained $f_\text{esc}^\text{CF} \sim 0.4$ for the Cosmic Horseshoe, but this disagrees with the upper limit of the absolute \fesc $<0.02$\ derived with HST imaging by \citet[][]{vasei}. However, \citet{quider} did not account for dust attenuation when deriving \fesc. In Sect.~\ref{horseshoe} we show that accounting for dust leads to \fesc\ values that are consistent with the HST observations of the Cosmic Horseshoe.

\subsection{Dust attenuation}
\label{dust}
The final contributor to the escape of ionizing photons in our fits is dust. Dust heavily impacts the observed stellar continuum at 912\AA: even small \ebv\ values lead to large attenuations. J1152$+$3400, with the smallest \ebv\ in the Lyman continuum emitting sample, has an A$_{912}$~=~1.7~mag ($\tau_\text{912} = 1.5$). Consequently, even small dust attenuation removes significant amounts of ionizing photons.  

The effect of dust is maximized in the idealistic case where there is only dust and no \hi\ along the line of sight (\cfhi~=1 and $\tau=0$). In this case, dust regulates the escape of ionizing photons. The contribution to the escape fraction solely from dust (\fescd) is calculated as
\begin{equation}
    f_\text{esc}^{D} = 10^{-0.4 E_\text{B-V} k_{912}}
,\end{equation}
where \fescd\ values are the closest to \fesco\ of the three parameters (cyan diamonds in Fig.~\ref{methods}). Nonetheless, \fescd\ is still too high to match \fesco, and the combination of dust and \cfhi\ are required to match the modeled \fesct\ (see purple triangles in Fig.~\ref{fig:methods}). 

The individual values of \ebv\ and \cfhi\ change depending on the assumed geometry \citepalias[Sect.~\ref{geometry};][]{gazagnes}. However, this does not diminish the contribution of either dust or \cfhi\ to \fesct. In an alternative geometry,  {the} clumpy geometry (Eq.~\ref{eq:clumpy}), the observed flux far from optically thick \hi\ lines (at wavelengths where $\tau_\lambda$ is small) is heavily influenced by the product of $ 10^{-0.4E_\text{B-V}k_\lambda} C_f^\text{H}$. Since most of the fitted wavelengths are actually in the small $\tau_\lambda$ regime, the attenuation significantly influences the fitted \cfhi\ value. While the exact contribution of dust and covering fraction are model dependent, \fesct\ depends on both.

\subsection{Connecting low attenuation to high-redshift leakers }
\label{lowext}
We find that dust attenuation strongly contributes to the predicted escape fractions. Consequently, low-mass--or equivalently low-metallicity--galaxies are ideal targets to emit ionizing photons. These properties are similar to the host galaxy properties of known local emitters \citep{Izotov16b,chisholm17}. Galaxies in the early universe should naturally have these properties \citep{bouwens12a, Madau14} and may have higher \fesc\ than local galaxies. \citet{schaerer10}  found that typical $<10^{10}$~M$_\odot$ galaxies at $z=6-8$ have A$_V < 1$. This implies that $f_\text{esc} > 0.05 (1-C_f^\text{H})$ for galaxies expected to reionize the universe. Using the median \cfhi\ from the Lyman continuum emitting sample (\cfhi~$=0.64$), $z=6-8$ galaxies should have $f_\text{esc} > 0.02$, much higher than the average galaxy at $z = 0$. Further, all of the $z\sim3-4$  confirmed LyC emitters have \ebv~$< 0.11$~mag, or $f_\text{esc} > 0.27(1-C_f^\text{H})$ \citep{deBarros16, shapley16, bian}. Using the median \cfhi\ from our Lyman continuum emitting sample, this corresponds to \fesc$>0.1$, which agrees with the \fesc\ required to reionize the universe at $z = 6-8$. Galaxies in the epoch of reioniziation likely have low dust attenuations, which makes them ideal candidates to emit a high fraction of their ionizing photons.

\section{Indirectly predicting the Lyman continuum escape fraction}
\label{predict}

\begin{figure*}
\includegraphics[width = \textwidth]{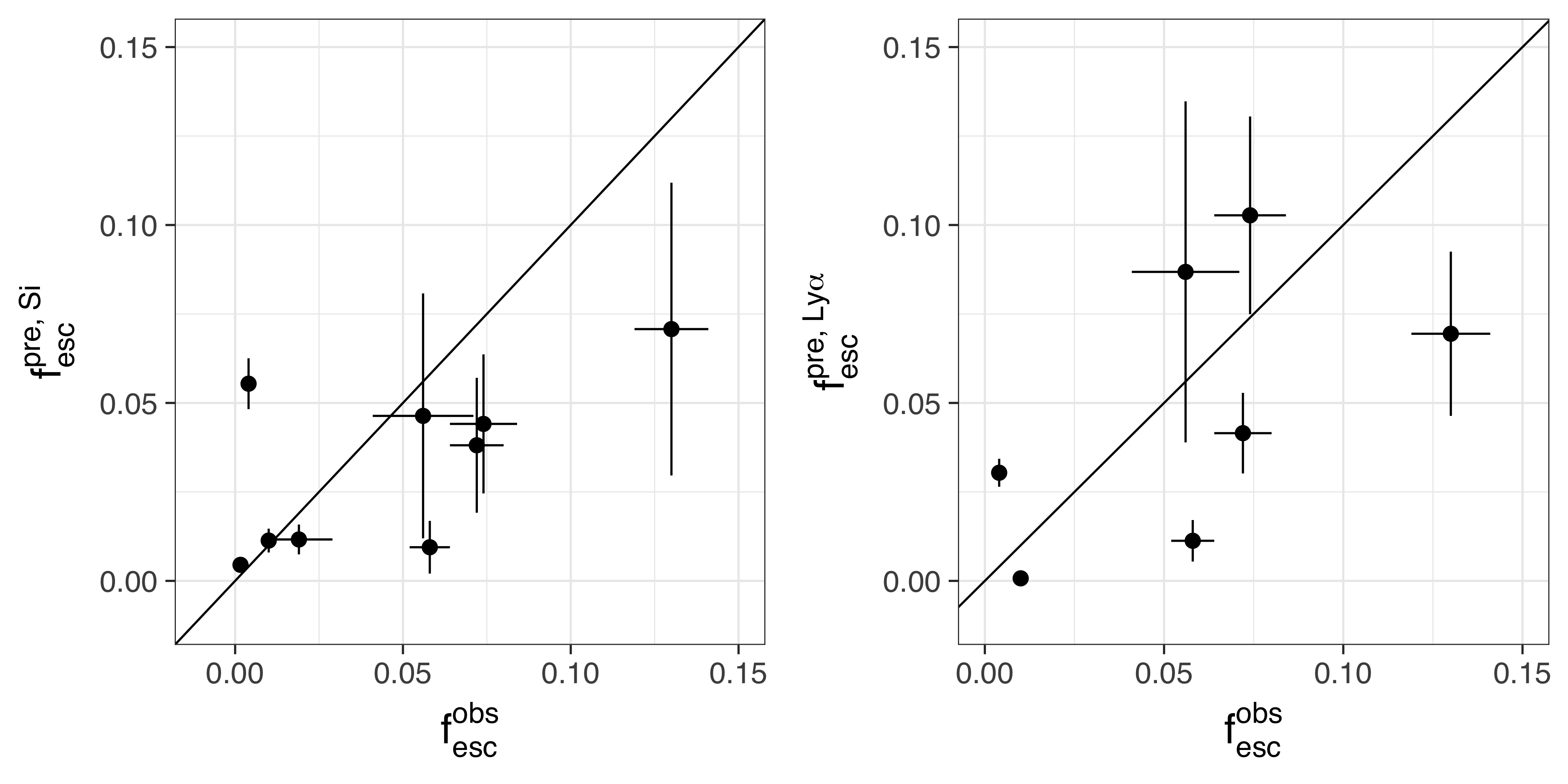}
\caption{\textit{Left panel:} Plot of the observed Lyman continuum escape fraction (\fesco) vs. the predicted Lyman continuum escape fractions made using the \siii covering fraction, derived from the \siiip~1260\AA\ and \siiip~1190\AA\ doublet. \textit{Right panel:} The escape fraction predicted by extinction correcting the Ly$\alpha$ escape fraction (\fesctla). The \fesctsi\ and \fesctla\ methods are consistent with the \fesco\ within 1.2 and 1.8$\sigma$, respectively. There are two fewer \fesctla\ points because \lya\ is in the detector gap for Tol~0440$-$381 and Mrk~54. }
\label{fig:predict}
\end{figure*}

Directly measuring \fesco\ requires deep rest-frame far-UV observations. This means that only a dozen galaxies are confirmed LyC emitters at any redshift. While the Lyman series accurately predicts the escape fraction (Fig.~\ref{fig:fesc}), the Lyman series is also not observable at high redshifts because the opacity of the circum-galactic medium is large. Therefore, we explored ancillary, indirect methods that can predict the \fesc\ of high-redshift galaxies. First we explored using the \siii covering fraction to predict the escape fraction. Then we used the observed \lya\ escape fraction to approximate \fesc. Finally, we used the ratio of the optical oxygen emission lines  (\ott=[\ion{O}{iii}]~5007\AA/[\ion{O}{ii}]~3727\AA). In Sect.~\ref{horse}, we illustrate how these three methods predict \fesct\ for galaxies that are not in our full sample because they do not have publicly available Lyman series observations. 

\subsection{Using Si II absorption}
\label{si2}

\siii has multiple absorption lines in the rest-frame far-UV, including the 1190 \AA\ doublet and 1260 \AA\ singlet. The ionization potential of \siii (16~eV) means that it probes partially neutral gas, and many studies have used it to diagnose LyC emitters \citep{Heckman11, Jones12, Jones13, Alexandroff15, chisholm17}. In \citetalias{gazagnes}, we showed that \cfhi\ and the \siii covering fraction (\cfsi) are linearly related, but not equal. We fit the relationship between \cfhi\ and \cfsi\ as\begin{equation}
    C_f^\text{pre, H} = \left(0.6 \pm 0.1\right)\times C_f^\text{Si} + \left(0.5 \pm 0.1\right) .
    \label{eq:cfh1si}
\end{equation}
This relationship is significant at the 3$\sigma$ significance level (p-value <0.001). This relation is statistically consistent with the relationship between \siiip~1260\AA\ and \hi\ found for $z\sim3$ galaxies in \citet{reddy}. In \citetalias{gazagnes}, we posited that this relation arises because metals do not completely trace the same gas as \hi, and \cfsi\ must be corrected to account for this differential covering. A multiple linear regression demonstrates that the constant in Eq.~\ref{eq:cfh1si} (0.6) depends on the gas-phase metallicity of the galaxy. This indicates that at lower metallicities the \siii traces a lower fraction of the \hi.

We predicted the escape fraction of ionizing photons using the \siii absorption lines as
\begin{equation}
f_\text{esc}^\text{pre, Si} = 10^{-0.4E_\text{B-V}k_{912}} (1-C_f^\text{pre, H}) ,
\label{eq:si2}
\end{equation}
where we used $k_{912} = 12.87$, the observed \ebv, and \cfhip\ from Eq.~\ref{eq:cfh1si}. The value \fesctsi\ is consistent with \fesco\ for the nine known Lyman continuum emitters (left panel of Fig.~\ref{fig:predict}). The difference between \fesctsi\ and \fesco\ is 46\% of the measured \fesco\ values. Similarly, the median \fesctsi\ is within 1.2$\sigma$ of \fesco. Using the \siii absorption predicts the observed escape fractions with similar accuracy as the Lyman series.

\subsection{Using Ly$\alpha$ escape fractions}
\label{lya}

Ionizing photons and \lya\ photons are related because \hi\ gas absorbs or scatters both \citep{verhamme15}. The \lya\ escape fraction is calculated as
\begin{equation}
    f_\text{esc}^{Ly\alpha} = \frac{\left(F[\text{Ly}\alpha]/F[\text{H}\alpha]\right)_\text{obs}}{\left(F[\text{Ly}\alpha]/F[\text{H}\alpha]\right)_\text{int}}
    \label{eq:fesclya}
,\end{equation}
where $\left(F[\text{Ly}\alpha]/F[\text{H}\alpha]\right)_\text{obs}$ is the observed ratio of the \lya\ flux to the extinction-corrected \ha\ flux, and  $\left(F[\text{Ly}\alpha]/F[\text{H}\alpha]\right)_\text{int}$ is the theoretical intrinsic flux ratio (which has a value of 8.7 for Case B recombination and a temperature of 10$^4$~K).  {The \fescly} measures the fraction of \lya\ photons that escape and does not directly depend on how the \lya\ photons escapes. Consequently, we assumed that the only difference between \fesc\ and \fescly is the dust attenuation, and used the \lya\ escape fraction to predict the LyC escape fraction (\fesctla) as
\begin{equation}
    f_\text{esc}^\text{pre, Ly$\alpha$} = 10^{-0.4E_\text{B-V}k_{912}} f_{esc}^\text{Ly$\alpha$} .
    \label{eq:lya}
\end{equation}
This implies that the LyC and \lya\ escape fractions are similar, but that the LyC escape fraction is lower because the dust attenuation is larger at 912\AA\ than at 1216\AA. Consequently, Eq.~\ref{eq:lya} effectively extinction corrects the \lya\ escape fraction to predict \fesc. These values are consistent with \fesco\ for the seven galaxies with measured \fescly (right panel of Fig.~\ref{fig:predict}). The average relative difference between \fesco\ and \fesctla\ is 55\% of \fesco, and \fesctla\ is, on average, within 1.8$\sigma$ of \fesco. The consistency of \fesctla\ is comparable to the two previous \fesct\ measurements. 

The similar \fesct\ and \fescly\ values are driven by the similar attenuations because the attenuation dominates \fesct\ (Sect.~\ref{methods}). The difference in calculating \fesctla\ and \fesct\ are the \cfhi\ and \fescly\ values (compare Eq.~\ref{eq:esc} and  Eq.~\ref{eq:lya}). This implies that \fescly\ and \cfhi\ are causally related \citep{dijkstra16,verhamme16}.

\subsection{Using \ott}
\label{o3}
\begin{figure}
\includegraphics[width = 0.5\textwidth]{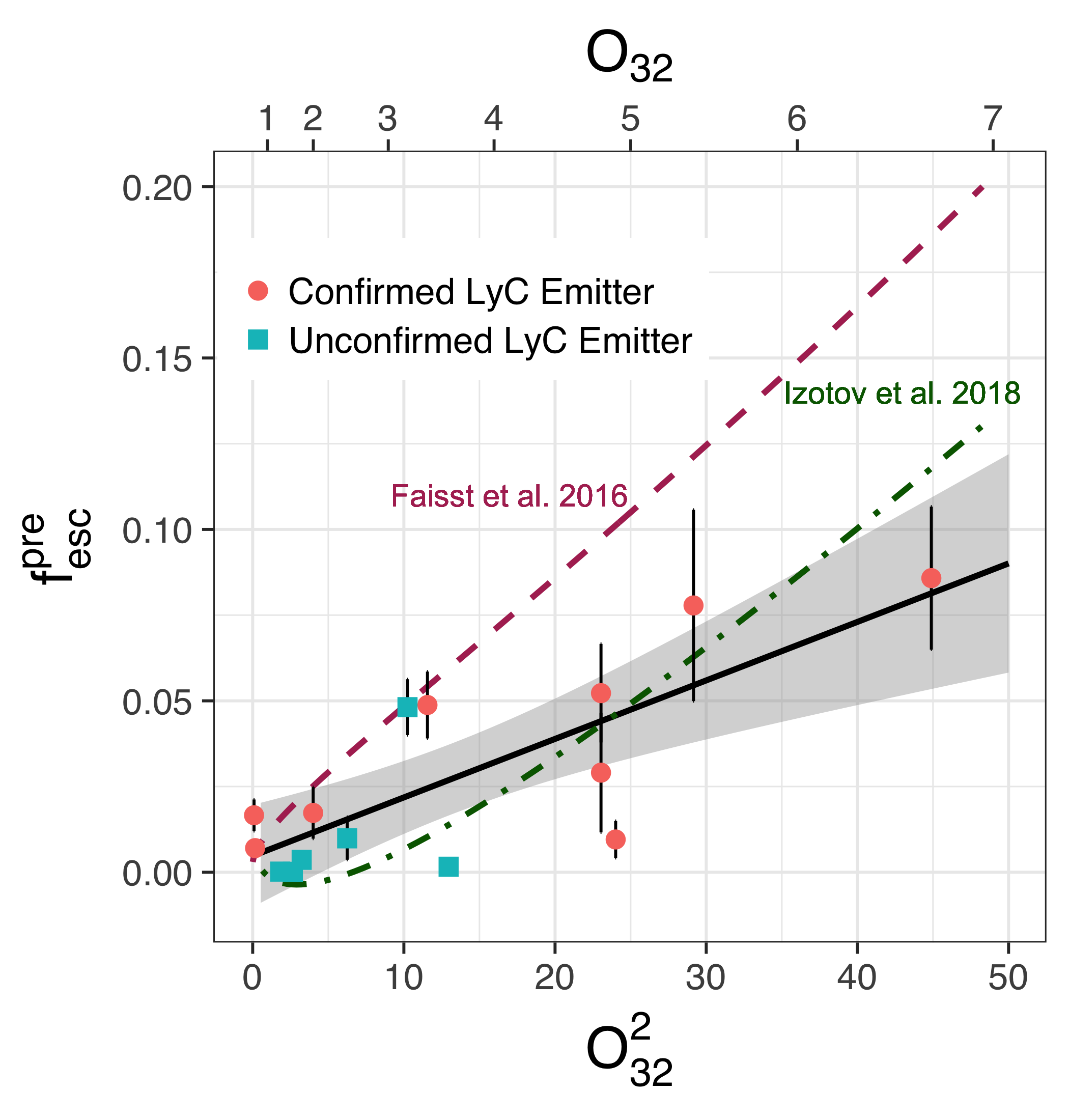}
\caption{Plot of the predicted Lyman continuum escape fraction (\fesct) from the Lyman series fits (Eq.~\ref{eq:esc}) vs. O$_{32}^2$ (\ott=[\oiiip~5007\AA]/[\oiip~3727\AA]) for the full sample from \citetalias{gazagnes}. The upper x-axis shows the corresponding linear \ott\ values. Red circles and blue squares denote confirmed and unconfirmed (i.e., galaxies without LyC observations) LyC emitters, respectively. The correlation (Eq.~\ref{eq:o32}) has a 3$\sigma$ significance (p-value < 0.001) and the 95\% confidence interval is shown in gray. Overplotted as a maroon dashed line is the empirical relationship from \citet{faisst}. The recent fit from \citet{izotov17} is also shown as the green dot-dashed line. The relationship derived here predicts lower \fesc\ values at large \ott\ than the two other relationships.}
\label{fig:o3}
\end{figure}

Historically, it was challenging to find galaxies emitting ionizing photons. A breakthrough came by selecting samples based on the [\ion{O}{iii}]~5007\AA/[\ion{O}{ii}]~3727\AA\ flux ratio (\ott), compactness, and large H$\beta$ equivalent widths. \citet{Izotov16}, \citet{Izotov16b}, and \citet{izotov17} found six out of six galaxies with \ott~>~4 had \fesco~>~0.05. This selection technique appears to efficiently select galaxies that emit ionizing photons based on their easily observed rest-frame optical properties. If this selection criteria is universally applicable, it is a powerful technique to select LyC emitting galaxies. It enabled \citet{faisst} to extend local \ott\ scaling relations to high redshifts to predict that $z > 6.5$ galaxies could reionize the universe.

\begin{table*}
\centering
\caption{Measured properties for the 7 galaxies from \citet{gazagnes} without observed Lyman continuum escape fractions.}
\begin{tabular}{lcccccc}
\hline
\hline
Galaxy Name & $z$ & \ebv\ & \cfhi\  & \fesct\ & \ott\ \\
 & &[mag] &  & [$\times10^{-3}$] & \\
\hline 
(1) & (2)  & (3) & (4) & (5) & (6)  \\
\hline 
J092600.4$+$442736\tablefootmark{a} & 0.18069& $0.11 \pm 0.01$ & $0.81\pm0.05$ & $50\pm10$ & 3.2  \\
GP~1244$+$0216\tablefootmark{b} &0.23942 & $0.29 \pm 0.04$ & $0.95 \pm 0.13$  & $2 \pm 1$ & 3.2 \\
GP~1054$+$5238\tablefootmark{b} & 0.25264 & $0.20 \pm 0.04$ & $0.89 \pm 0.16$ & $10 \pm 4$ & 2.5 \\
GP~0911$+$1831\tablefootmark{b} & 0.26223 & $0.35 \pm 0.04$ & $0.77 \pm 0.12$  & $4 \pm 2$ & 1.8\\ 
SGAS~J152745.1$+$065219\tablefootmark{c} & 2.7628 & $0.37 \pm 0.002$ & $0.99 \pm 0.04$  & $0.1 \pm 0.010$ & 1.6 \\
SGAS~J122651.3$+$215220\tablefootmark{c} & 2.9260 & $0.20 \pm 0.001$ & $1.00 \pm 0.01$  & $0.35 \pm 0.01$ & 1.4 \\
\hline
 GP~0303$-$0759\tablefootmark{b} & 0.16488 & $0.12 \pm 0.05$ &  --  &  --  &7.3 \\
 J142947.00$+$064334.9\tablefootmark{a} & 0.1736 & $0.11 \pm 0.02$ & $0.96 \pm 0.06$  & $10 \pm 1$ & - \\
 The Cosmic Eye\tablefootmark{c} & 3.0748 & $0.41 \pm 0.01$ & $1.00\pm 0.02$ & $0.016\pm0.0005$ & - & \\ 
\hline
\end{tabular}
\tablefoot{Column 1 gives the galaxy name  listed in descending \ott\ order. Column 2 gives the redshifts of the galaxies. Column 3 is the stellar attenuation (\ebv) measured using the stellar continuum fitting of \citetalias{gazagnes}. Column 4 is the \hi\ covering fraction measured from the depths of the Lyman series lines (\cfhi). Column 5 is the predicted Lyman continuum escape (\fesct) calculated using the Lyman series absorption properties. The sixth column gives the [\ion{O}{iii}]~5007\AA/[\ion{O}{ii}]~3727\AA\ flux ratio (\ott).  We note that GP~0303$-$0759, J142947.00$+$064334.9, and the Cosmic Eye (the three galaxies below the horizontal line) are not included in Sect.~\ref{o3} because GP~0303$-$0759 does not have a measured \cfhi\ owing to a Milky Way absorption line, and J142947.00$+$064334.9 and the Cosmic Eye do not have literature \ott\ values.} 
\tablebib{(a) \citet{Heckman11, Heckman15, Alexandroff15, Heckman16} (b) \citet{Henry15} (c) \citet{Wuyts, rigby17a}}
\label{tab:full}
\end{table*}

To test the effect of \ott\ on the ionizing escape fraction, we used the full sample of 15 galaxies with predicted \fesct\ using the Lyman series (Eq.~\ref{eq:esc})  and \ott\ measurements from \citetalias{gazagnes}; the Cosmic Eye and J1429$+$0643 are excluded because they do not have measured \ott , and GP~0303$-$0759 is excluded due to Milky Way contamination. By including these  six  galaxies, with unobserved LyC emission, we extended the \ott\ dynamic range and derived a relationship between \ott\ and \fesct\ (Fig.~\ref{fig:o3}).

We first explored whether \ott\ scales with \fesct. We tested a variety of models for the scaling of the two variables:  linearly, quadratically, or as a logarithm of each (or both) variable. We maximized the F-statistic for a model where the variables scale as  \fesct-$\text{O}_{32}^{2}$. This relationship is significant at the  3$\sigma$ significance (p-value < 0.001; R$^2$ =  0.61; Fig.~\ref{fig:o3}). A linear regression (see the line in Fig.~\ref{fig:o3}, with the shaded 95\% confidence region) gives a relationship of
 \begin{equation}
    f^\text{pre, O}_\text{esc} = \left(0.0017 \pm 0.0004\right)\text{O}_{32}^2 + \left(0.005 \pm 0.007\right) .
    \label{eq:o32}
\end{equation}
This predicts \fesc\ using easily observed rest-frame optical emission lines.

Fig.~\ref{fig:o3} also shows the empirical relationship from \citet{faisst}. The two relations are discrepant at \ott\ values corresponding to  \fesc~$ > 0.05$. Eq.~\ref{eq:o32} predicts that more than $10$\% of the ionizing photons escape galaxies when \ott~$ > 5.7$. Using the extrapolation of \ott\ with redshift from \citet{faisst}, the average galaxy does not have \fesc~$ = 0.1$  until $z \sim 11$.  $z \sim 11$ is marginally consistent with the $z_\text{re} = 8.8^{+1.7}_{-1.4}$ redshift of instantaneous reionization derived from the combination of the Planck lensing and polarization studies \citep{planck}.

Fig~\ref{fig:o3} also compares Eq.~\ref{eq:o32} to a similar trend found by \citet{izotov17}. These authors used a recently discovered galaxy, J1154$+$2443 with an exceptionally high \fesco~$=0.46$ to derive a relationship between \ott\ and \fesco\ (the dot-dashed green curve in Fig.~\ref{fig:o3}). Many of our \fesct\ values agree with the \citet{izotov17} relation and the two relationships are consistent for \fesct\ values up to \fesc~$\sim0.1$. However, the \citet{izotov17} relationship increases more rapidly at higher \ott\ and \fesct\ values than Eq.~\ref{eq:o32} does. This is apparent from the galaxy J1154$+$2443, which has \ott~$=11.5\pm 1$. The expected \fescto, $0.26\pm0.06$, is nearly 3$\sigma$ lower than \fesco. This suggests that Eq.~\ref{eq:o32} may steepen at larger \ott, but the steep portion of the \citet{izotov17} trend is largely driven by the one high \fesco\ galaxy. If Eq.~\ref{eq:o32} steepens at higher \ott\ then the redshift required for galaxies to emit 10\% of their ionizing photons would be lower than $z \sim 11$. Further observations, probing a uniform and large range of \ott, are required to refine Eq.~\ref{eq:o32}.

Studies often use low \nhi\ values to explain the correlation between \fesc\ and \ott\ \citep[so-called "density-bounded" regimes;][]{Jaskot13, zackrisson13, Nakajima13}. However, \ott\ arises both from high ionization parameter (as required in the density-bounded regime) and from low metallicities \citep[][]{nagao, Nakajima14, Shapley15, Sanders, chisholm17, strom}. As shown in \citetalias{gazagnes} and Fig.~\ref{fig:methods}, LyC photons escape because the \hi\ covering fraction and dust attenuation are low, not because the \hi\ column density is low. Rather, the low attenuation likely connects \ott\ and \fesc. Low attenuation could be related to high ionization parameters (dust is destroyed) and/or low metallicities (dust is not created). In this scenario, the lower dust content could mean that there are not enough metals to uniformly fill the gas, or that there are not enough metals to efficiently cool the gas. Both result in channels with little dust or \hi\ along the line of sight, allowing for more ionizing photons to escape the galaxy. We find a $2\sigma$ trend between \ott\ and \ebv\ in our sample. Thus, the correlation between \fesc\ and \ott\ may reflect the low dust attenuation of LyC emitters. However, further observations, spanning a large range of \ott, and more theoretical work are required to confirm and understand this observed correlation.

\subsection{Using  multiple methods to predict \fesc}
\label{multiple}
Above, we demonstrated that the \siii absorption lines, \fesclyp, and \ott\ consistently predict \fesco,  but the \ott\ method needs further data to verify. These three prediction methods have similar deviations from \fesco\ as using the \hi\ absorption lines. However, the individual methods do not always precisely reproduce \fesco. If the LyC cannot be directly observed, then \fesct\ should be calculated using as many of the three methods as possible. The mean and standard deviation of the three different methods then approximates \fesct. As an example, Tol~1247$-$232 has a largely discrepant \fesct, but when the average of \fesctla, \fesctsi, and \fescto\ are  taken \fesct$ = 0.038\pm0.022$, which is consistent, within 2$\sigma$, with \fesco. Estimating \fesc\ with multiple methods reduces the systematic errors of individual observations and produces more consistent \fesc\ predictions. We illustrate this in the next section with observations of both high- and low-redshift galaxies.

It is important to produce a statistical sample of predicted \fesc\ values with all of the different methods to determine the systematics of each method. For instance, direct observations of the LyC, as well as \fesc\ inferred from the Lyman series, \lya, and \siii\ lines are all line-of-sight geometry dependent estimates of \fesc, such that the inferred value substantially changes whether the orientation is through a hole in the \hi\ or through an \hi\ cloud. Conversely, the \ott\ ratio depends less on the geometry because the ISM is relatively optically thin to the [\oiiip] and [\oiip] emission. These effects may be imprinted on the different predicted \fesc\ values, and variations in \fesct\ may illustrate physical geometric variations in LyC emitting galaxies. 

\section{Predicting the Lyman continuum escape fraction of galaxies without Lyman series observations}
\label{horse}

\begin{figure}
\includegraphics[width = 0.5\textwidth]{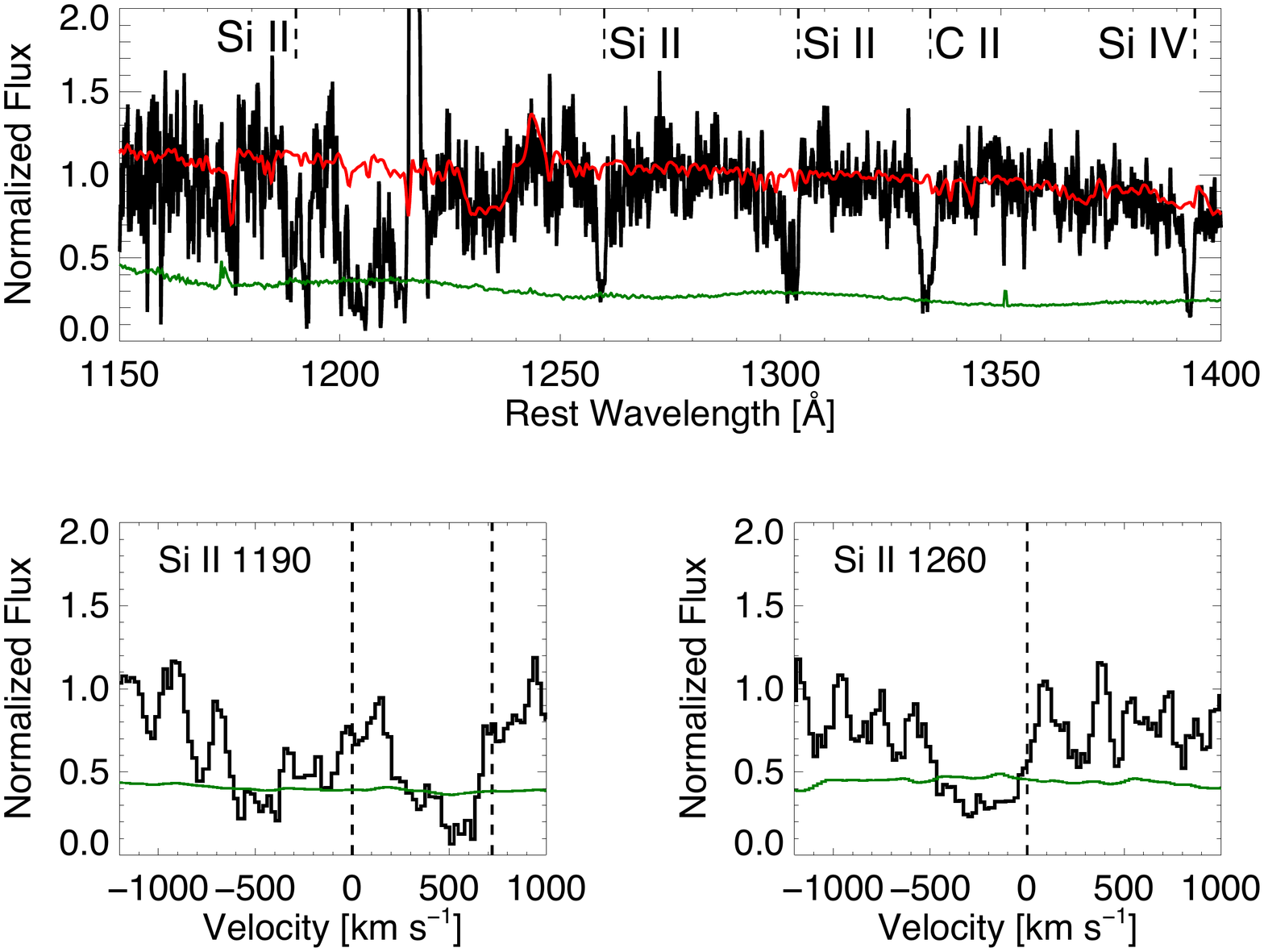}
\caption{\textit{Top panel: } Rest-frame UV spectra between $1150-1400$\AA\ of the Cosmic Horseshoe, a $z = 2.38$ gravitationally lensed galaxy from the \megasaura\ sample \citep{rigby17a}. Overplotted in red is the best-fit {\small STARBURST99} stellar continuum fit. This fit measures \ebv~$= 0.16$~mag. The error spectrum is included underneath in dark green. {\it Bottom panels:} The  \siiip~1190\AA\ doublet (left) and \siiip~1260\AA\ singlet (right). The corresponding \cfsi\ from the \siiip~1260\AA\ line is 0.77. Vertical dashed lines indicate the zero velocity of the various strong ISM metal absorption lines (labeled in the upper panel).}
\label{fig:horse}
\end{figure}

The relations presented in the previous section enable estimation of \fesc, even if the Lyman continuum or Lyman series are not observable. This is especially important for $z>4$ galaxies because the IGM transmission of the LyC is $<38$\% at $z>4$ \citep{songaila}, making LyC detections even more challenging. The three indirect probes in the previous section may be the only way to estimate the emissivity of high-redshift galaxies reionizing the universe (Eq.~\ref{eq:emis}). We test the methods of Sect.~\ref{predict} by fitting the rest-frame UV spectra between $1200-1500$\AA\ of a few test cases in the same manner as we did in Sect.~\ref{predict}. These test cases are the Cosmic Horseshoe, the \megasaura\ sample, Haro~11, a recently discovered strong LyC emitter from \citet{izotov17}, and high-redshift confirmed LyC emitters.  Because of the uncertainty of the \ott\ relation (see Sect.~\ref{o3}), we only comment on what the observed \ott\ values imply for \fesct. Table~\ref{tab:predicted} lists the parameters used to predict the escape fractions for each galaxy.

\begin{table*}
\centering
\caption{Predicted Lyman continuum escape fractions for the 8 galaxies with predicted escape fractions higher than 0.01, but without Lyman series or Lyman continuum observations.}
\begin{tabular}{lcccccccc}
\hline
\hline
Galaxy Name & $z$ & \ebv & \cfsi & \fesctsi & \fesctla & \fesct & \fesco \\
 &  & [mag] &  & &  &  &  \\
\hline 
(1) & (2)  & (3) & (4) & (5) & (6) & (7) & (8)\\
\hline 
\textit{Ion2} & 3.212 & <0.04 & - &- & >0.49 & >0.49 & $0.64^{+1.1}_{-0.1}$\tablefootmark{a} \\
SDSS~J1154+2443 & 0.3690 & 0.06 & - & - & 0.48 &  0.48 & $0.46 \pm 0.02$\tablefootmark{b} \\
SGAS~J211118.9$-$011431 & 2.8577 & 0.12 & 0.30 & 0.082 & - &  0.082 & - \\
SGAS~J142954.9$-$120239 & 2.8245 & 0.08 & 0.40 & 0.080 & - &  0.080 & - \\
Haro~11 & 0.0206 & 0.12 & 0.60 & 0.036 & -& 0.036 & $0.033\pm0.007$\tablefootmark{c} \\
SGAS~J090003.3$+$223408 & 2.0326 & 0.11 & 0.65 & 0.026 & 0.025 & $0.026\pm0.001$ & $0.015 \pm 0.012$\tablefootmark{d}\\ 
SGAS~J095738.7$+$050929 & 1.8204 & 0.21 & 0.63 & 0.013 & - &0.013 & - \\
SGAS~J145836.1$-$002358 & 3.4868 & 0.07 & 0.83 & 0.011 & -  & 0.011 & - \\
The Cosmic Horseshoe &2.3812 & 0.16 & 0.77 & 0.009 & 0.012 &  $0.011\pm0.002$ & $<0.02$\tablefootmark{e} \\
\hline
\end{tabular}
\tablefoot{Column 1 gives the galaxy name. Column 2 gives the redshifts of the galaxies. Column 3 and 4 give the stellar attenuation (\ebv) and \siii covering fraction (\cfsi) determined from a stellar continuum fit similar to the methods detailed in Sect.~\ref{fit}. Column 5 gives the Lyman continuum escape fraction predicted using the \siiip~1260\AA\ absorption line (\fesctsi; Eq.~\ref{eq:si2}). Column 6 gives the Lyman continuum escape fraction predicted by extinction correcting the Ly$\alpha$ escape fraction (\fesctla; Eq.~\ref{eq:lya}).  Column 7 gives the mean and standard deviation of the predicted Lyman continuum escape fractions. Column 8 gives the observed Lyman continuum escape fraction (\fesco). The table is ordered in descending \fesct. All of the galaxies, except \textit{Ion2}, Haro~11, and J1154+2443 are drawn from the \megasaura\ sample \citep[Sect.~\ref{meg};][]{rigby17a}. Unmeasured quantities are denoted with dashes.}
\tablebib{
(a) \citet{deBarros16}; (b) \citet{izotov17}; (c) \citet{Leitet13}; (d) This work (Sect.~\ref{j09}); (e) \citet{vasei}}
\label{tab:predicted}
\end{table*}

\subsection{ {The} Cosmic Horseshoe}
\label{horseshoe}

The Cosmic Horseshoe \citep{Belokurov} is an ideal test case for these methods. At $z = 2.38$, it is one of the best-studied gravitationally lensed galaxies. However, from the methods presented in Sect.~\ref{predict}, we would not expect the Cosmic Horseshoe to strongly emit ionizing photons. Restframe UV spectra from the \megasaura\ sample \citep[Fig.~\ref{fig:horse};][]{rigby17a} show a young stellar population with relatively deep \siii absorption lines (i.e., large \cfsi). Similarly, \lya\ observations from the Echellette Spectrograph and Imager on the KECK~II telescope only find \fescly$=0.08$ \citep{quider}. Moreover, the Cosmic Horseshoe has a relatively small extinction-corrected \ott\ \citep[2;][]{hainline}. The suspicions of low \fesc\ are confirmed by deep HST LyC imaging that measures an upper limit of the absolute escape fraction of \fesco~$<0.02$ \citep{vasei}. \citet{vasei} noted that there was a 20\% chance that the low \fesco\ arises from IGM attenuation.  {While the IGM attenuation  has a low-probability of impacting the \fesco,} proper simulations of the IGM opacity can quantify the impact of the IGM opacity at higher redshifts \citep{shapley16}.

From the stellar continuum fit in Fig.~\ref{fig:horse} (red line), we measured an \ebv\ of 0.16~mag, consistent with \citet{quider}. The \siiip~1260\AA\ profile has a \cfsi\ of 0.77 (corresponding to a \hi\ covering fraction of \cfhip~$=0.94$ using Eq.~\ref{eq:cfh1si}). The escape fraction predicted using \cfsi\ and Eq.~\ref{eq:si2} is \fesctsi~$= 0.009$. The measured \fescly~$=0.08$ leads to a LyC escape fraction of \fesctla~$=0.012$ (Eq.~\ref{eq:fesclya}).  Finally, the extinction-corrected \ott\ is small, such that Eq.~\ref{eq:o32} implies a low \fescto~=0.011 (Eq.~\ref{eq:o32}). 

 Combining the two robust estimates of \fesct\ (the \siii and \lya\ values), we derive a mean estimate of \fesct~$=
0.011\pm0.002$ (Table~\ref{tab:J0900}). This result is also consistent with the escape fraction predicted by the \ott\ scaling relation. The \fesct\ satisfies the upper limit of the absolute escape \fesco$<0.02$ from the HST imaging \citep{vasei}.

\subsection{Other \megasaura\ galaxies}
\label{meg}
The Cosmic Horseshoe is one of 14 galaxies within the \megasaura\ sample \citep[three are included in the full sample in Table~\ref{tab:full};][]{rigby17a}. We also predicted the escape fraction for the full \megasaura\ sample. Fitting the stellar continua and \siiip~1260\AA\ covering fractions of the sample predicts that only six (42\%) have \fesctsi~$>0.01$ (Table~\ref{tab:predicted} gives the predicted escape fractions of these six galaxies). Two \megasaura\ galaxies with low \fesctsi~$<0.01$, SGAS~J1226$+$2152 and SGAS~J1527$+$0652, also have  {low} \ott\ values of 1.4 and 1.6, respectively \citep[Table~\ref{tab:full};][]{Wuyts}.  These  {low} \ott\ values correspond to \fescto~$< 0.01$, consistent with their low \fesctsi. No \megasaura\ galaxy has \fesctsi~$> 0.1$. This is consistent with the nondetection of LyC photons in the individual spectra \citep{rigby17a}. 

\subsubsection{SGAS~J0900$+$2234}
\label{j09}
\begin{table}
\centering
\caption{Observed properties of SGAS~J0900$+$2234}
\begin{tabular}{ccc}
\hline
\hline
Row & Property & Value\\
\hline 
(1) & $F[H\alpha]$ & $(225\pm 56) \times 10^{-17}$\\
(2) & $F[Ly\alpha]$ & $(185\pm15) \times 10^{-17}$\\
(3) & \fescly & $0.09\pm0.02$ \\
(4) & $F[1500]_\text{obs}$ & $(6.9 \pm 0.4) \times 10^{-18}$ \\
(5) & $F[900]_\text{obs}$ & $(1.8 \pm 1.4) \times 10^{-19}$ \\
(6) & $(F[1500]/F[900])_\text{int}$ & 1.4 \\
(7) & \ebv\ & $0.11\pm0.001$ \\
\hline
\end{tabular}
\tablefoot{Row 1 gives the extinction-corrected H$\alpha$ flux \citep{bian2010}. Row 2 gives the observed \lya\ flux. Row 3 gives the measured \lya\ escape fraction. Row 4 gives the measured flux at 1500\AA\ from the HST F475W image. Row 5 gives the measured flux at 912\AA\ from the HST 218W image \citep{bian}. Row 6 gives the ratio of the flux at 1500\AA\ and 912\AA\ from the stellar population fit. Row 7 gives the measured attenuation (\ebv; in mags) from the stellar population fit. All flux values have units of erg~s$^{-1}$~cm$^{-2}$. }
\label{tab:J0900}
\end{table}

SGAS~J0900$+$2234, a $z = 2.03$ lensed galaxy, is the second best \megasaura\ test case. First, combining the rest-frame optical observations from \citet{bian2010} with the \megasaura\ data estimates the \lya\ escape fraction to be 0.09 (Table~\ref{tab:J0900}). This \fescly\ leads to \fesctla~$=0.025$, consistent with the \fesctsi~$=0.026$ (Table~\ref{tab:predicted}). There are no literature [\ion{O}{ii}]~3727\AA\ observations for this galaxy. Consequently, we predicted \fesct~$= 0.026\pm0.001$.  

This lensed galaxy has both LyC \citep[F218W; rest-frame central wavelength of 734\AA; PID: 13349; PI: X. Fan; ][]{bian} and rest-frame FUV (F475W; rest-frame central wavelength of 1566\AA; PID: 11602; PI: S. Allam) HST imaging. \citet{bian} have not detected significant LyC photons from this lensed galaxy, but the HST images provide weak constraints on \fesco. Following Eq.~1 from \citet{Leitet13}, we estimated \fesco\ as
\begin{equation}
    f^\text{obs}_{esc} = \frac{(F[1500]/F[900])_\text{int}}{(F[1500]/F[900])_\text{obs}} 10^{-0.4 E_\text{B-V} k_{1500}}
    \label{eq:frel}
\end{equation}
where we took $k_{1500}$ from the \citet{reddy_ext} attenuation law and \ebv\ from the {\small STARBURST99} stellar continuum fit. The intrinsic flux ratio is  measured from the {\small STARBURST99} stellar population fit to the spectra (Table~\ref{tab:J0900}), and is similar to values from \citet{Izotov16b} for an instantaneous 7~Myr stellar population (the fitted stellar age). $(F[1500]/F[900])_\text{obs}$ is the observed ratio of the flux at 1500\AA\ and 900\AA, respectively. The $F[1500]$ and $F[900]$ values are measured from the exact same regions in the F475W and F218 images.  $F[900]$ has a low significance (1.3$\sigma$; Table~\ref{tab:J0900}), which led \citet{bian} to not report a significant LyC detection. We measured \fesco~$ = 0.015 \pm 0.012$, consistent, within 1$\sigma$, with \fesct\ (Table~\ref{tab:predicted}). 

\subsection{Haro 11}
\label{haro}

Haro~11, a nearby star-forming galaxy has a measured \fesco~$= 0. 033 \pm 0.007$ from Far-Ultraviolet Spectroscopic Explorer (FUSE) observations \citep{Bergvall06, leitet11}. A recent HST/COS spectrum of Knot C in Haro~11 covers rest-frame 1130-1760\AA\ \citep{Heckman11, Alexandroff15}. We measured \ebv~$=0.124$~mags and \cfsi~$=0.60$ from this COS spectrum \citep{Chisholm16}. This \cfsi\ value agrees with the recent value from \citet{rivera17} and leads to \fesctsi~$=0.036$. \citet{keenan} have measured the \ott\ of Haro~11 using HST/WFC3 imaging. While the imaging makes it challenging to robustly subtract the stellar continuum, the \ott\ ratio is between 2--4 for Knot C.  This corresponds to a \fescto~$= 0.01-0.03$ (Eq.~\ref{eq:o32}). Both \fesctsi\ and \fescto\ are broadly consistent, within 1$\sigma$, with \fesco.  

\subsection{J1154$+$2443}
\label{J1154}
\citet{izotov17} have recently discovered a new low-redshift LyC emitter, J1154$+$2443, with HST/COS spectra. At \fesco~$=0.46\pm0.02$, it has the highest observed escape fraction in the local universe. \citet{izotov17} have listed properties of J1154$+$2443 that nicely align with those that we suggest lead to a high \fesco: low metallicity ($\oh = 7.65$), low extinction ($A_\text{v} = 0.145$, or \ebv~$=0.06$ using their $R_\text{v}= 2.4$), high \fescly (\fescly~$=0.98$), and large \ott\ (\ott~$=11.5$). 
 Using the \fescly\ and converting the attenuation measured with the \citet{Cardelli89} curve to an attenuation using the \citet{reddy_ext} relation, we predict that J1154+2443 would have \fesctla~$=0.48$ (Table~\ref{tab:predicted}). This is consistent with \fesco\ found by \citet{izotov17}.  In Sect.~\ref{o3} we found the \ott\ relation under-predicts the \fescto\ of this galaxy by $3\sigma$. This suggests that more observations are required to better constrain the \ott\ relation at large \ott.
 
\subsection{High-redshift LyC emitters}

While most of the confirmed LyC detections have come at low redshift ($z < 0.4$), four $z \sim 3-4$ galaxies have confirmed \fesco\ \citep{Vanzella15, deBarros16, shapley16, bian, vanzella17}. These galaxies are typically more extreme LyC emitters than the $z \sim 0$ galaxies (\fesco~$=0.2-0.7$) and have characteristics that Sect.~\ref{predict} suggests lead to high \fesco: low \ebv, weak \siii\ absorption lines, and strong \lya. \textit{Ion2} \citep{Vanzella15} is the only high-redshift galaxy with literature limits for \ott\ or \fesclyp, while no high-redshift galaxy has a published \cfsi. \textit{Ion2} has an extreme \fesco~$= 0.64^{+1.1}_{-0.1}$, an upper limit of \ott$>15$, a very low dust extinction upper limit (\ebv~$<0.04$~mag), a large lower limit of \fescly~$> 0.78$, and a nondetected \siiip~1260\AA\ absorption line \citep{deBarros16}. Using the methods in Sect.~\ref{predict} we predict \fescto$>0.39$ and \fesctla~$>0.49$. These predicted lower limits of the LyC escape fraction are consistent with \fesco~$=0.64$ from \citet{deBarros16}.

\subsection{Prospects for the epoch of reionization}
\label{eor}
The above examples indicate that the methods from Sect.~\ref{predict} can powerfully predict the \fesc\ of high-redshift galaxies. {\it JWST} and ELTs will observe the rest-frame UV of field and lensed $z=6-8$ galaxies to estimate \fesctsi. Additionally, optical emission lines will be redshifted to 3.5--4.5~$\mu$m, such that the NIRSpec instrument on {\it JWST} will measure H$\alpha$ and \ott. These observations will estimate \fesctla\ and \fescto. Combined, the three methods predict \fesc\ values that are broadly consistent with the observed escape fractions of local Lyman continuum emitting galaxies. The escape fractions, the total number of ionizing photons, and the luminosity functions will then describe whether star-forming galaxies reionized the $z=6-8$ universe.

\section{Summary}
\label{summary}
We analyzed the rest-frame UV spectra of nine low-redshift ($z<0.3$) star-forming galaxies that emit ionizing photons. In a companion paper \citep{gazagnes}, we fit the stellar continuum, dust attenuation, Lyman series absorption lines (\hi\ absorption lines blueward of \lya), and ISM metal absorption lines. Here, we combined the \hi\ column densities and covering fractions with the dust attenuations to predict the fraction of ionizing photons that escape local galaxies. The Lyman continuum and Lyman series both directly trace the escape of ionizing photon, but neither are observable at redshifts greater than 4. Therefore, we tested three indirect ways of estimating \fesc: the \siii absorption lines (Sect.~\ref{si2}),  {the} Ly$\alpha$ escape fraction (Sect.~\ref{lya}), and  {the} [\ion{O}{iii}]/[\ion{O}{ii}] flux ratio (Sect.~\ref{o3}). We then used these methods to predict the escape fractions of galaxies without Lyman series observations to illustrate how these indirect methods can estimate the escape fraction of high-redshift galaxies (Sect.~\ref{horse}).

The major results of this study are as follows:
\begin{enumerate}
    \item The radiative transfer equation (Eq.~\ref{eq:esc}), along with the fits to the dust attenuation, \hi\ covering fraction, and \hi\ column density reproduce the observed \fesc\ to within 1.4$\sigma$ (Fig.~\ref{fig:fesc}). The Lyman series absorption properties accurately predict the escape fraction of ionizing photons.
    \item As shown in \citet{gazagnes}, the observed \hi\ column densities indicate that the Lyman continuum is optically thick. Instead, ionizing photons escape because the covering fraction is less than one (Fig.~\ref{fig:methods}).
    \item The covering fraction alone overpredicts the escape fraction. While geometry dependent (see the discussion in Sect~\ref{geometry}), dust attenuation is a key ingredient for the escape of ionizing photons  (Fig.~\ref{fig:methods}). Estimating the escape fraction as (1-\cf) will overestimate the true escape fraction. 
    \item Indirect methods also provide accurate estimates of the escape fraction of ionizing photons. The \siii absorption line and extinction-correction Ly$\alpha$ escape fraction predicts \fesc\ with similar accuracy as the Lyman series (Fig.~\ref{fig:predict}), while the square of the [\ion{O}{iii}]/[\ion{O}{ii}] flux ratio scales strongly with \fesc\ (3$\sigma$ significance; Fig.~\ref{fig:o3}). The [\ion{O}{iii}]/[\ion{O}{ii}]  relation agrees with previous studies for low [\ion{O}{iii}]/[\ion{O}{ii}] values, but underpredicts the \fesc\ of higher [\ion{O}{iii}]/[\ion{O}{ii}] values. This suggests that a larger sample of large [\ion{O}{iii}]/[\ion{O}{ii}] galaxies is required to constrain the full trend. 
    \item We applied the indirect methods to galaxies without Lyman series observations to illustrate how these methods predict \fesc\ (Table~\ref{tab:predicted}). In all cases, the \fesc\ values predicted with indirect methods are consistent with either the observed \fesc\ or upper limits of \fesc. Most (58\%) of the $z = 1.7-3.6$ \megasaura\ galaxies have low escape fractions (\fesc~$\le 1$\%), while the remaining galaxies have predicted  \fesc~$\sim 1-8$\% (Sect.~\ref{meg}). Additionally,  the predicted escape fraction of  J1154$+$2443, a recently discovered local galaxy with a high escape fraction (0.46), agrees with the observed value. Overall, the predicted \fesc\ values are consistent with the observed escape fractions.
\end{enumerate}
This analysis presents new methods to measure and analyze the escape fractions of galaxies in the epoch of reionization using {\it JWST} and future ELT(Sect.~\ref{eor}). Deep rest-frame UV and optical spectroscopy of these high-redshift galaxies may determine whether high-redshift galaxies emit sufficient ionizing photons to reionize the universe or if other sources are required.

\bibliographystyle{aa} 
\bibliography{references} 

\end{document}